\documentclass[10pt,a4paper]{article}

\usepackage[utf8]{inputenc}
\usepackage[T1]{fontenc}
\usepackage{lmodern}
\usepackage[a4paper,left=4.9cm,right=2.0cm,top=2.3cm,bottom=2.6cm,
            marginparwidth=3.6cm,marginparsep=0.6cm,headheight=30pt]{geometry}
\usepackage{amsmath,amssymb}
\usepackage{graphicx}
\usepackage[dvipsnames]{xcolor}
\usepackage{fancyhdr}
\usepackage{titlesec}
\usepackage{enumitem}
\usepackage{booktabs}
\usepackage{listings}
\usepackage[font=small,labelfont=bf,labelsep=period]{caption}
\usepackage[round,sort&compress]{natbib}
\usepackage{microtype}
\usepackage[colorlinks=true,linkcolor=MidnightBlue,citecolor=MidnightBlue,
            urlcolor=MidnightBlue,breaklinks=true]{hyperref}
\usepackage{url}
\usepackage[section]{placeins}

\newcommand{\flake}{FLAKE}
\newcommand{\code}[1]{\texttt{#1}}
\newcommand{\eps}{\epsilon}
\newcommand{\rcm}{\mathbf{r}_\mathrm{cm}}

\definecolor{sidegray}{gray}{0.35}
\definecolor{codebg}{gray}{0.96}

\titleformat{\section}{\sffamily\Large\bfseries}{}{0pt}{}
\titleformat{\subsection}{\sffamily\large\bfseries}{}{0pt}{}
\titleformat{\paragraph}[runin]{\sffamily\bfseries}{}{0pt}{}[.]
\titlespacing*{\section}{0pt}{2.0ex plus 0.5ex}{1.0ex}
\titlespacing*{\subsection}{0pt}{1.6ex plus 0.4ex}{0.6ex}

\setlist{itemsep=0.15em,topsep=0.2em}
\reversemarginpar

\newcommand{\shorttitle}{\flake: rigid 2D clusters on crystalline substrates}
\fancyfootoffset[L]{\dimexpr\marginparwidth+\marginparsep\relax}
\fancypagestyle{first}{%
  \fancyhf{}%
  \fancyheadoffset[L]{\dimexpr\marginparwidth+\marginparsep\relax}%
  \fancyhead[L]{\includegraphics[height=0.9cm]{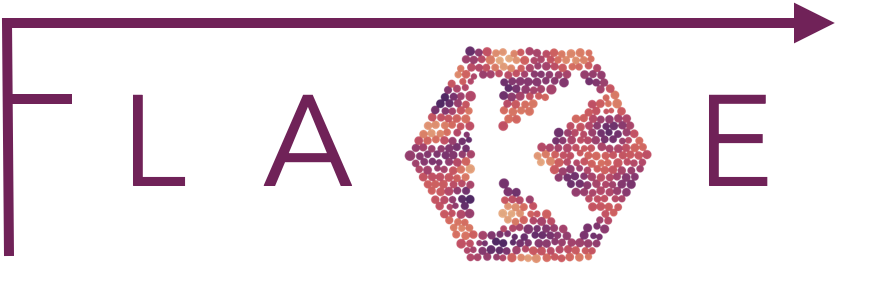}}%
  \fancyhead[R]{\footnotesize\sffamily\color{sidegray}\today}%
  \fancyfoot[L]{\footnotesize\sffamily\color{sidegray}Silva (2026). \shorttitle.}%
  \fancyfoot[R]{\footnotesize\sffamily\thepage}%
  }

\newcommand{\sidehead}[1]{{\bfseries #1}\par}
\newcommand{\sidebar}{%
  \marginpar{\raggedright\sffamily\scriptsize\color{sidegray}
  \sidehead{DOI}
  [waiting for JOSS]\par\medskip
  \sidehead{Software}
  \href{https://github.com/LamaKing/flake_rigid}{Repository} (GitHub)\\
  \href{https://flake-rigid.readthedocs.io}{Documentation} (Read the Docs)\\
  \href{https://pypi.org/project/flake-rigid/}{Package} (PyPI: \code{flake-rigid})\par\medskip
  \sidehead{Version}
  0.1.3\par\medskip
  \sidehead{Contact}
  \href{mailto:ansilva@sissa.it}{ansilva@sissa.it}\par}}

\title{\sffamily\bfseries FLAKE: Friction and Lattice Analysis of Kinetics and
Energetics of rigid 2D clusters on crystalline substrates}
\author{\sffamily Andrea Silva\,\href{https://orcid.org/0000-0001-6699-8115}{%
  \textsuperscript{\normalfont ORCID}}\,\textsuperscript{1,2}\\[0.6em]
  \small\normalfont
  \textsuperscript{1}\,SISSA -- Scuola Internazionale Superiore di Studi Avanzati, Trieste, Italy
  (ROR \href{https://ror.org/004fze387}{004fze387})\\
  \small\normalfont
  \textsuperscript{2}\,CNR-IOM, Trieste, Italy
  (ROR \href{https://ror.org/00yfw2296}{00yfw2296})}
\date{}

\hypersetup{pdftitle={FLAKE: Friction and Lattice Analysis of Kinetics and
  Energetics of rigid 2D clusters on crystalline substrates},
  pdfauthor={Andrea Silva},
  pdfkeywords={python, condensed matter physics, friction, superlubricity,
  nanotribology, Langevin dynamics, moire pattern}}

\begin{document}
\makeatletter
\begin{flushleft}
  \vspace*{-0.8em}
  {\LARGE\@title\par}
  \vspace{0.8em}
  {\large\@author\par}
  \vspace{0.4em}
  {\small\sffamily\color{sidegray}\textbf{Keywords:} python, condensed matter
  physics, friction, superlubricity, nanotribology, Langevin dynamics, moir\'e
  pattern\par}
\end{flushleft}
\makeatother
\thispagestyle{first}

\section{Summary}\label{sec:summary}
\sidebar
When a small solid island slides across a crystalline surface, the outcome (smooth flow or stick-slip jerking, free rotation or angular locking) is
controlled almost entirely by the geometric relationship between the island and
the surface: their relative orientation, size, and lattice symmetry. These
phenomena are central to nanoscale friction and to the interpretation of
colloidal monolayer experiments \citep{cao_orientational_2019,
cao_pervasive_2021} and are the premises for the study of \emph{structural
superlubricity} in real interfaces \citep{wang_colloquium_2024,
hirano_superlubricity_1990, dienwiebel_superlubricity_2004,
ying_scalingup_2025, yao_superlubricity_2025}.

\flake{} (Friction and Lattice Analysis of Kinetics and Energetics) is a Python
package for simulating the statics and Langevin dynamics of a rigid 2D cluster
sliding over a periodic or quasiperiodic substrate potential. It is designed for
researchers in nanotribology and 2D-materials physics who need fast, systematic
exploration of how geometry and substrate symmetry control
the energy landscape and friction, i.e. studies that are prohibitively expensive
with fully atomistic simulations. \flake{} has been the primary simulation tool
for three published studies on structural superlubricity and colloidal friction
\citep{silva_frictionless_2023, cao_moire_2022, liao_twisting_2023}.

\section{Statement of need}\label{sec:need}
Nanotribology experiments with colloidal clusters
\citep{cao_orientational_2019, cao_pervasive_2021, cao_moire_2022} and van der
Waals heterostructures \citep{liao_twisting_2023, yao_superlubricity_2025}
show phenomena (orientational locking, rotational depinning,
coupled roto-translational motion) that are difficult to interpret without a
corresponding simulation tool. The standard computational approach in this
field is all-atom molecular dynamics with realistic interatomic potentials,
typically using LAMMPS \citep{thompson_lammps_2022} and ASE \citep{larsen_ase_2017}. While indispensable for
elastic effects and chemical specificity, all-atom simulations are
prohibitively expensive for the systematic parameter sweeps needed to extract
scaling laws and build phase diagrams. They also entangle elastic deformation
with the purely geometric commensurability effects that dominate in stiff 2D
materials.

Several recent computational studies address
the phenomena \flake{} targets: rotational energy landscapes and
scaling laws for twisted bilayers \citep{zhang_universal_2025,
zhang_frictionless_prl_2025, zhang_directional_2026}, shape-dependent friction
scaling \citep{yan_shape_2024, gao_frictional_2025}, orientational control via
strain \citep{zhou_orientational_2025}, and experimental evidence for
directional superlubricity in van der Waals heterostructures
\citep{lester_directional_2024}. These studies use a variety of tools and
conventions, often relying on LAMMPS with material-specific force fields that
constrain the accessible geometries and substrate symmetries. Reviews of the
scaling-up challenge \citep{ying_scalingup_2025} highlight that systematic
parameter exploration remains a bottleneck. \flake{} is a geometry- and
chemistry-agnostic Python package for fast, systematic rigid-body friction
studies, and a baseline for elastic LAMMPS studies.

\section{State of the field}\label{sec:field}
Structural superlubricity in 2D materials is an active research front
\citep{wang_colloquium_2024, ying_scalingup_2025}. The geometric, rigid-body
mechanism is now well-established in principle \citep{silva_frictionless_2023},
but quantitative predictions (e.g.\ how friction scales with cluster size,
shape and orientation) still require
systematic numerical exploration. No existing open Python package targets this
combination for the rigid-body tribology use case, so each group writes its
own simulation code.

\paragraph{DFT-level PES tools}
TribChem \citep{losi_tribchem_2023} and the high-throughput workflow of
\citet{wolloch_highthroughput_2022} operate at the opposite length-scale
extreme: they compute interfacial adhesion and shear strength from
first-principles electronic structure for periodic unit-cell-sized systems.
These tools are complementary to \flake{}: their output
(corrugation amplitude $\eps$ and well shape) can parametrise \flake's substrate
potentials for specific material interfaces.

\paragraph{Geometric descriptors}
The Registry Index \citep{hod_registry_2013} characterises interfacial
commensurability from atomic geometry alone, producing qualitative maps.
However, it carries no energy scale and cannot drive dynamics or compute energy
barriers.

\paragraph{Interlayer potentials}
In LAMMPS simulations of 2D-material interfaces, the interlayer interaction is
typically described by the Interlayer Potential (ILP) of
\citet{leven_ilp_2016}, which uses Gaussian-shaped atom--atom overlap terms for
registry-dependent repulsion, qualitatively captured by \flake's
Gaussian-well substrate.

\section{Software design}\label{sec:design}

\subsection{Rigid-body reduction}
The central design choice in \flake{} is the rigid-body reduction. In 2D, a cluster of
$N$ particles on a substrate in principle requires $2N$ equations of
motion. Treating the cluster as a rigid body collapses this to three collective
degrees of freedom (center-of-mass position $\mathbf{r}=(x,y)$ and
orientation angle $\theta$) governed by
\begin{align}
  \eta_t\,\dot{\mathbf{r}} &= \mathbf{F}_\mathrm{sub}(\mathbf{r},\theta)
      + \mathbf{F}_\mathrm{ext} + \boldsymbol{\xi}_t, \label{eq:eom-t}\\
  \eta_r\,\dot{\theta} &= \tau_\mathrm{sub}(\mathbf{r},\theta)
      + \tau_\mathrm{ext} + \xi_r, \label{eq:eom-r}
\end{align}
where $\eta_t = N\eta$ and $\eta_r = \eta\sum_i r_i^2$ are translational and
rotational drag coefficients computed from cluster geometry,
$\mathbf{F}_\mathrm{sub}$ and $\tau_\mathrm{sub}$ are the total substrate force
and torque summed over all particles, and $\boldsymbol{\xi}_{t,r}$ are Gaussian
white noise terms,
$\langle \xi_i(t)\xi_j(t')\rangle = 2k_BT\eta_i\,\delta_{ij}\,\delta(t-t')$.
The overdamped (inertia-free) limit is the physically appropriate regime for
colloidal clusters in viscous media \citep{cao_orientational_2019} and for 2D
materials at low sliding speeds; extending the integrator to include inertia is
straightforward. This reduction makes it possible to run thousands of
trajectories in minutes on a laptop, enabling the systematic parameter sweeps
over cluster size, shape, orientation, substrate symmetry, and applied
force that are the primary target of \flake.

Rigidity is a good approximation for stiff 2D materials
(graphene, hBN, MoS$_2$) at moderate loads, and for colloidal clusters where
interparticle distances are effectively fixed: in both cases the dominant friction mechanism
is geometric commensurability rather than elastic deformation
\citep{silva_frictionless_2023}.

\paragraph{Limitations}
The rigid-body approximation does not capture intralayer elasticity, which
drives the Aubry-type commensurate--incommensurate transition
\citep{brazda_aubry_2018, wang2025aubry}; moir\'e-induced out-of-plane buckling in freestanding
bilayers \citep{wang_bending_2023, wang_buckling_2024}; or elastic edge effects that give rise to
chiral moir\'e dynamics at finite contacts \citep{silva_chirality_2024}. These
effects require fully elastic simulations and are beyond \flake's
scope.

\subsection{Performance}
All substrate
energy and force kernels are Numba just-in-time compiled (\code{@njit}, Numba's
JIT decorator \citep{lam_numba_2015}): the \code{substrate\_from\_params}
factory pre-converts parameters to float64 arrays once and returns closures,
eliminating per-step Python overhead, and \code{run\_md} dispatches to a fully
JIT-compiled inner loop when no Python callbacks are needed; the test suite
checks that this loop is faster than the interpreted one.

\subsection{String method in roto-translational space}
The string method for minimum-energy paths \citep{weinan_string_2002} is
extended to the joint $(x,y,\theta)$ space using independent length scales
$(l_x,l_y,l_\theta)$ that put translational and angular displacements on equal
footing,
\begin{equation}
  ds^2 = \left(\frac{dx}{l_x}\right)^2 + \left(\frac{dy}{l_y}\right)^2
       + \left(\frac{d\theta}{l_\theta}\right)^2 ,
\end{equation}
a metric choice required to find physically meaningful
roto-translational paths. The scales enter only the reparametrisation of the
string to equal arc length, not the gradient step; typical choices are
$l_x=l_y$ equal to the substrate spacing and $l_\theta$ equal to the angular
period of the contact (e.g.\ $60^\circ$ for triangular-on-triangular).

\subsection{Features}
\flake{} provides:
\begin{itemize}
  \item \emph{Energy landscape maps} over translational and/or rotational
    degrees of freedom (\code{flake map}), for static analysis of
    commensurability and orientational locking.
  \item \emph{Minimum-energy path search} via the string method in 2D
    (translational) and 3D (roto-translational) configuration space
    (\code{flake string}), giving static friction forces and transition
    pathways.
  \item \emph{Parameter-sweep Langevin dynamics} (\code{flake sweep})
    parallelised over a parameter grid via joblib, for depinning thresholds and
    velocity--force scaling laws.
  \item \emph{Substrate types}: sinusoidal (optical lattices), Gaussian well
    (ILP-like registry dependence), tanh well (lithographic patterns), and flat
    (testing baseline); new types require only a single Numba \code{@njit}
    function.
  \item \emph{Cluster geometries}: circle, hexagon, triangle, rectangle,
    ellipse, and arbitrary polygon via Shapely; multi-site bases supported.
  \item A \emph{YAML-driven CLI}, which needs no Python scripting, and six end-to-end Jupyter notebooks in
    \code{examples/} covering the full workflow from substrate construction to
    scaling-law analysis.
\end{itemize}

\section{Research impact}\label{sec:impact}
\flake{} v0.1.3 is the packaged and tested release of a codebase
developed from 2022 onward at SISSA. The predecessor (v0.0.6, publicly
available since November 2022) was the primary simulation tool for the
following published studies:
\begin{itemize}
  \item \citet{silva_frictionless_2023} (Nanoscale): geometric classification
    of arbitrary 2D interfaces and scaling laws of static friction force from
    finite commensurate and incommensurate contacts.
  \item \citet{cao_moire_2022} (Physical Review X): coupled roto-translational
    depinning and the role of moir\'e pattern evolution in colloidal
    experiments.
  \item \citet{liao_twisting_2023} (ACS Applied Materials and Interfaces):
    combined experiment and simulation of twisting dynamics in large
    lattice-mismatch van der Waals heterostructures.
\end{itemize}
The colloidal experiments of \citet{cao_orientational_2019} and
\citet{cao_pervasive_2021} provided the physics foundation and motivated the
initial design of the package.

Version 0.1.3 adds a fully JIT-compiled MD inner loop, the roto-translational
string method, and a complete test suite. The package is currently in active
use at SISSA for ongoing research on quasicrystalline substrates and
roto-translational synchronisation effects, in collaboration with
experimentalists in Germany and China.

\paragraph{Availability}
\flake{} requires Python $\geq3.10$, NumPy, SciPy, Numba, PyYAML, h5py and
joblib (Shapely and ASE are optional) and is installed with
\code{pip install flake-rigid}. Its pytest suite, run in continuous
integration, includes physics validation tests: $F_c=NF_{1s}$ for commensurate
clusters and reduced static friction for incommensurate ones.
\FloatBarrier

\section{AI usage disclosure}\label{sec:ai}
The physical model, all design decisions, and the research direction of
\flake{} are entirely the work of the author. The central argument that
geometry dominates friction in stiff 2D contacts, making a rigid-body model
both sufficient and necessary for systematic study is grounded in the
author's prior publications \citep{silva_frictionless_2023, cao_moire_2022}.
The choice of substrate models, the roto-translational string method and its
arc-length metric, and the scope of the package all predate any AI
involvement: the core physical model and simulation algorithms were functional
in the predecessor codebase (v0.0.6, November 2022) prior to any AI
assistance. Claude Code (Anthropic, \code{claude-sonnet-4-6}) was subsequently
used to improve the codebase: making the code more readable (docstring
rewriting, module restructuring and API cleanup) and more efficient (Numba JIT
refactoring of substrate closures and the MD inner loop), adding the test
suite and CI configuration, and improving this manuscript. All AI-generated outputs were reviewed and edited by
the author.

\section*{Acknowledgements}
The author thanks Emanuele Panizon, Jin Wang, Dong Han, Erio Tosatti, Andrea
Vanossi, Nicola Manini, Xin Cao, and Clemens Bechinger for the physics
discussions and experimental collaborations that motivated and shaped this
work. The author thanks Andrea Trost for help with the logo design.

This work was supported by the European Research Council (ERC) under the
European Union's Horizon 2020 research and innovation programme, grant
agreement No.~834402 (ULTRADISS), and by the Italian Ministry of University
and Research (MUR) under PRIN grant No.~20178PZCB5 (UTFROM).

\bibliographystyle{plainnat}
\bibliography{refs}

\clearpage \newpage

\appendix

\setcounter{figure}{0} \setcounter{table}{0}
\renewcommand{\thefigure}{S\arabic{figure}}
\renewcommand{\thetable}{S\arabic{table}}
\providecommand{\theHfigure}{}\renewcommand{\theHfigure}{S\arabic{figure}} \providecommand{\theHtable}{}\renewcommand{\theHtable}{S\arabic{table}}
\renewcommand{\thelstlisting}{S\arabic{lstlisting}}

\section{Appendix: Workflow and examples}\label{sec:workflow}
A typical study builds the cluster, defines the substrate, scans the static
energy landscape, and drives the cluster with Langevin dynamics. All
quantities are in reduced units: well depth $\eps=1$, substrate spacing $a=1$,
single-particle drag $\eta=1$.

\subsection{Building clusters}
\begin{figure}[!htb]
  \centering
  \includegraphics[width=\linewidth]{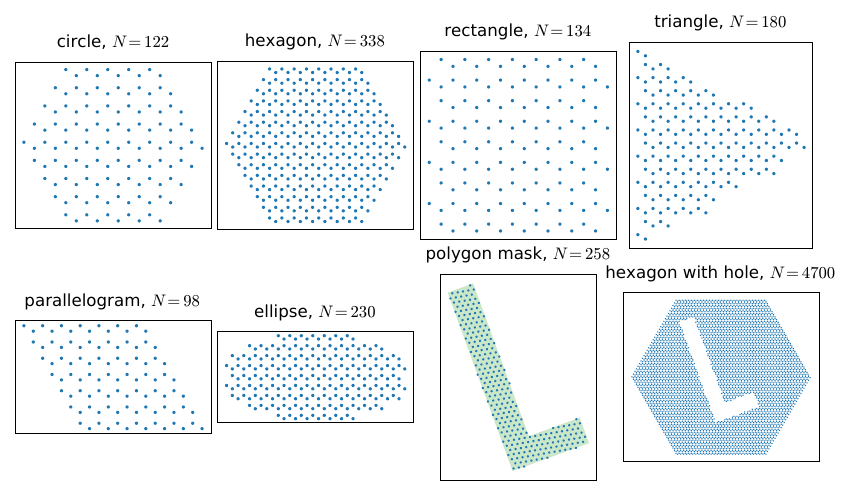}
  \caption{Cluster geometries: the six built-in shapes cut from a triangular
  lattice with a two-site (honeycomb) basis; an L-shaped polygon (shaded) masking
  a triangular lattice rotated independently of it; the same polygon as a hole
  in a honeycomb hexagon.}
  \label{fig:clusters}
\end{figure}

A cluster is a finite cut of a 2D Bravais lattice, optionally decorated with a
multi-site basis (Fig.~\ref{fig:clusters}). \code{make\_cluster} provides the
built-in shapes; \code{cluster\_poly} masks the lattice with an arbitrary
Shapely polygon, rotated independently of the lattice, or punches it as a hole;
\code{params\_from\_poscar} imports the in-plane lattice and basis of a real
material from a VASP POSCAR file. Positions are returned with the center of mass
at the origin and $\theta=0$: rotation and translation are initial conditions,
applied by the caller. The drag coefficients of
Eqs.~\eqref{eq:eom-t}--\eqref{eq:eom-r} follow from the same positions.

\subsection{Substrates}
\begin{figure}[!htb]
  \centering
  \includegraphics[width=\linewidth]{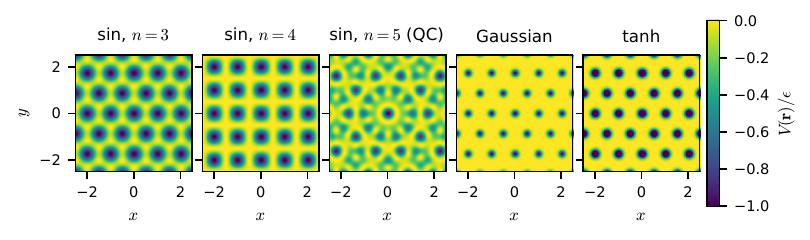}
  \caption{Single-particle substrate potential for the available families:
  sinusoidal with $n=3$, $4$ and $5$ (quasicrystal, QC) plane waves, and
  triangular lattices of Gaussian and tanh wells.}
  \label{fig:substrates}
\end{figure}

The substrate is a fixed potential acting on each particle; force and torque on
the cluster are sums over particles, with the torque taken about the center of
mass. \emph{Sinusoidal} substrates superpose $n$ plane waves
\citep{vanossi_static_2012},
\begin{equation}
  V(\mathbf{r}) = -\frac{\eps}{n^2}\Bigl|\sum_{l=0}^{n-1}
      e^{i\mathbf{k}_l\cdot\mathbf{r}}\Bigr|^2 ,
\end{equation}
with $n=2,3,4$ giving lines, triangular and square lattices, and $n=5,6$
quasicrystals with no unit cell. \emph{Lattice-of-wells} substrates repeat a
Gaussian well, smoothly cut off at a finite radius, or a flat-bottomed tanh well
as used for lithographic colloidal templates \citep{cao_pervasive_2021}
(Fig.~\ref{fig:substrates}). A flat substrate serves as a testing baseline.
\code{substrate\_from\_params} turns a parameter dictionary or YAML file into
JIT-compiled closures for energy, force and torque.

\subsection{Static analysis}
\begin{figure}[!htb]
  \centering
  \includegraphics[width=\linewidth]{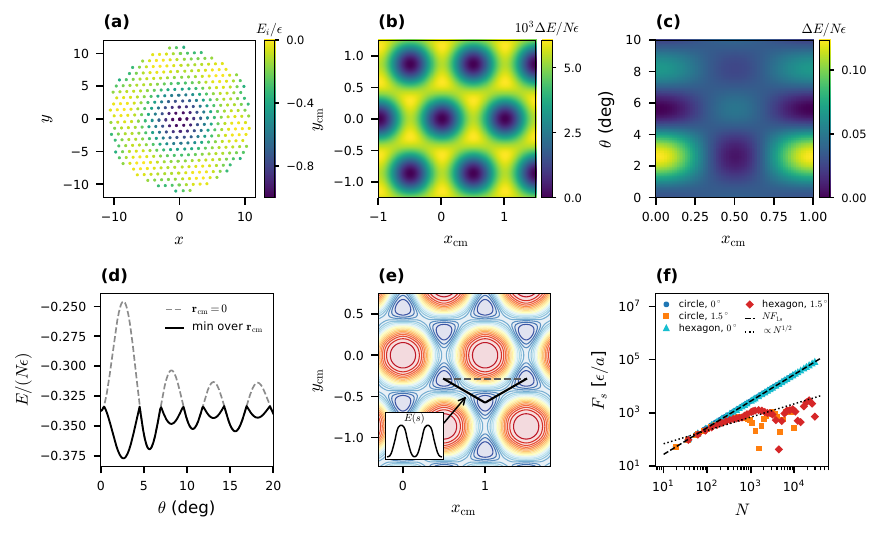}
  \caption{Static analysis. (a)--(d): circular cluster ($N=397$, mismatch
  $\rho=1.05$) on the triangular sinusoidal substrate. (a)~Particle energies at
  $\theta=2.5^\circ$, showing the moir\'e pattern. (b)~Translational map at
  $\theta=0$. (c)~Energy versus $x_\mathrm{cm}$ and $\theta$ at
  $y_\mathrm{cm}=0$. (d)~Orientational landscape: energy minimised over the CM
  position (solid) and at fixed $\rcm=0$ (dashed). (e)~String method for a
  different system: a circular cluster ($N=637$, $\rho=1+1/21$, $\theta=0$) on a
  triangular lattice of tanh wells; initial string (dashed) and
  relaxed minimum-energy path (solid) on the translational landscape (red: high,
  blue: low); inset: energy along the path. (f)~Static friction versus $N$ for
  circular and hexagonal clusters, aligned ($0^\circ$) and twisted
  ($1.5^\circ$); dashed: $NF_{1s}$; dotted: $N^{1/2}$ guide to the eye. In (b)
  and (c) energies are per particle, measured from the minimum.}
  \label{fig:landscapes}
\end{figure}

\code{translational\_map}, \code{rotational\_map} and \code{rototrasl\_map}
evaluate energy, force and torque on grids of CM positions, angles, or both
(Fig.~\ref{fig:landscapes}). Panels (a)--(d) show a circular cluster with a
5\% lattice mismatch on the triangular sinusoidal substrate. The particle
energies trace the moir\'e pattern [panel (a)], and the cancellation of particle
forces makes the translational corrugation per particle more than two orders
of magnitude smaller than for a single particle [panel (b)]. Translation and
rotation are coupled: the translational corrugation depends strongly on
$\theta$ [panel (c)]. Minimising the energy over the CM position at each angle
gives the orientational landscape, whose global minimum lies at a small twist,
$\theta\approx2.6^\circ$ [panel (d)].

\code{find\_mep} locates minimum-energy paths with the string method, in
$(x_\mathrm{cm},y_\mathrm{cm})$ at fixed orientation or in the full
$(x_\mathrm{cm},y_\mathrm{cm},\theta)$ space. The barrier is the energy maximum
along the path and the static friction is the largest force met,
$F_s=\max_s|\mathbf{F}(s)|$, with $\mathbf{F}=-\nabla E$ the force along the
path $s$. In panel (e), on a lattice of tanh wells, the straight
initial string between two equivalent minima relaxes to a path that bends
through an intermediate minimum and crosses two equivalent saddles.

Repeating such scans over cluster size gives scaling laws [panel (f)]: a
commensurate contact follows $F_s=NF_{1s}$ exactly, with $F_{1s}$ the
single-particle value, while a small twist makes the contact incommensurate and
$F_s$ sublinear, with oscillations set by the fraction of moir\'e tiles in the
cluster \citep{silva_frictionless_2023}.

\FloatBarrier

\subsection{Driven dynamics and depinning}
\begin{figure}[!htb]
  \centering
  \includegraphics[width=\linewidth]{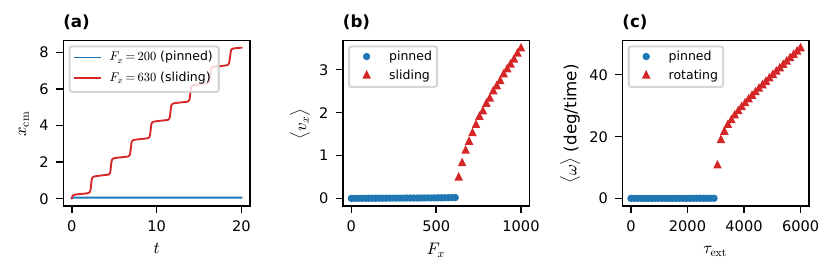}
  \caption{Overdamped Langevin dynamics of a commensurate circular cluster
  ($N=223$) on the triangular sinusoidal substrate. (a)~CM trajectory below
  (pinned) and above (sliding) the depinning threshold. (b)~Drift velocity
  versus applied force. (c)~Mean angular velocity versus applied torque.}
  \label{fig:dynamics}
\end{figure}

\code{run\_md} integrates Eqs.~\eqref{eq:eom-t}--\eqref{eq:eom-r} with the
Euler--Maruyama scheme, with noise fixed by the fluctuation--dissipation
theorem. \code{sweep\_md} runs one trajectory per point of a parameter grid in
parallel, reduces each with a post-processing function (e.g.\ drift velocity),
and optionally stores trajectories in HDF5 with resume support.
Figure~\ref{fig:dynamics}(a) shows a commensurate cluster pinned below and
sliding above the depinning threshold; panels (b) and (c) show translational
and rotational depinning.

\FloatBarrier

\subsection{Usage}
Listing~\ref{lst:api} builds the commensurate cluster of
Fig.~\ref{fig:dynamics} and runs an energy evaluation, a rotational map, a
translational minimum-energy path, whose static friction matches the depinning
force of Fig.~\ref{fig:dynamics}(b), and a driven trajectory above threshold.
From the shell, the physical parameters sit in a YAML file and each task reads
its own configuration, writing HDF5 output (examples in \code{example\_cli/}):
\begin{lstlisting}[language=bash]
flake map    -i params.yaml --grid grid_trasl.yaml -o map_trasl.h5
flake string -i params.yaml --cfg  string_roto.yaml -o mep_roto.h5
flake sweep  -i params.yaml --spec sweep_Fx.yaml
\end{lstlisting}

\begin{lstlisting}[caption={Minimal use of the Python API.},label={lst:api},captionpos=b]
import numpy as np
from flake.substrate import substrate_from_params, get_ks
from flake.cluster import cluster_from_params, rotate
from flake.maps import rotational_map
from flake.string_method import find_mep
from flake.dynamics import run_md

params = {
    # substrate: triangular sinusoidal potential of depth epsilon
    'well_shape': 'sin', 'epsilon': 1.0, 'sub_basis': [[0., 0.]],
    'ks': get_ks(1.0, 3, 4./3., 0.),     # R, n, c_n, alpha_n (radians)
    # cluster: circular cut of a commensurate triangular lattice
    'cluster_shape': 'circle', 'N1': 15, 'N2': 15, 'cl_basis': [[0., 0.]],
    'a1': [1., 0.], 'a2': [0.5, -np.sqrt(3.)/2.],
}
pen_func, en_func, _ = substrate_from_params(params)  # JIT-compiled closures
pos = cluster_from_params(params)       # (N, 2), CM at origin, theta = 0

cm = np.zeros(2)
E, F, tau = en_func(rotate(pos, 2.0) + cm, cm)      # theta = 2 deg
rot = rotational_map(pos, en_func, theta_deg=np.linspace(-30, 30, 601))
mep = find_mep(pos, en_func, p0=[0., 0.], p1=[1., 0.], n_pt=50)
traj = run_md(pos, en_func, eta=1.0, Fx=630., kBT=1e-8,
              dt=1e-3, n_steps=20000)
\end{lstlisting}

\end{document}